\RequirePackage{ifpdf}
\ifpdf 
\documentclass[pdftex]{sigma}
\else
\documentclass{sigma}
\fi

\usepackage{bbm}

\def\a{\alpha}
\def\b{\beta}
\def\g{\gamma}

\def\de{\delta}
\def\eps{\epsilon}

\def\th{\theta}

\def\la{\lambda}
\def\m{\mu}
\def\n{\nu}

\def\vp{\varphi}

\newcommand{\C}{\mathbb C}
\newcommand{\R}{\mathbb R}
\newcommand{\Z}{\mathbb Z}
\newcommand{\unity}{\mathbbm{1}}

\def\im{\textrm{i}}
\def\ep{\textrm{e}}
\def\diff{\textrm{d}}
\def\tr{\textrm{tr}}
\def\pa{\partial}

\def\>{\rangle}
\def\<{\langle}
\def\+{\dagger}
\def\Tdag{T^{\dagger}}
\def\zb{{\bar z}}
\def\dn{a^2}
\def\up{a^{\+2}}
\def\gi{\frac1G}

\begin{document}

\allowdisplaybreaks

\renewcommand{\thefootnote}{$\star$}

\renewcommand{\PaperNumber}{045}

\FirstPageHeading

\ShortArticleName{The Noncommutative Ward Metric}

\ArticleName{The Noncommutative Ward Metric\footnote{This paper is a
contribution to the Special Issue ``Noncommutative Spaces and Fields''. The
full collection is available at
\href{http://www.emis.de/journals/SIGMA/noncommutative.html}{http://www.emis.de/journals/SIGMA/noncommutative.html}}}

\Author{Olaf LECHTENFELD~$^{\dag\ddag}$ and Marco MACEDA~$^\S$}

\AuthorNameForHeading{O. Lechtenfeld and M. Maceda}

\Address{$^\dag$~Institut f\"ur Theoretische Physik,
Leibniz Universit\"at Hannover, \\
\hphantom{$^\dag$}~Appelstra\ss{}e 2, 30167 Hannover, Germany}
\EmailD{\href{mailto:lechtenf@itp.uni-hannover.de}{lechtenf@itp.uni-hannover.de}}
\URLaddressD{\url{http://www.itp.uni-hannover.de/~lechtenf/}}

\Address{$^\ddag$~Centre for Quantum Engineering and Space-Time Research,
Leibniz Universit\"at Hannover, \\
\hphantom{$^\ddag$}~Welfengarten 1, 30167 Hannover, Germany}

\Address{$^\S$~Departamento de Fisica, UAM-Iztapalapa, \\
\hphantom{$^\S$}~A.P. 55-534, C.P. 09340, M\'exico D.F., M\'exico}
\EmailD{\href{mailto:mmac@xanum.uam.mx}{mmac@xanum.uam.mx}}

\ArticleDates{Received January 31, 2010, in f\/inal form May 27, 2010;  Published online June 02, 2010}

\Abstract{We analyze the moduli-space metric in the static non-Abelian charge-two
sector of the Moyal-deformed $\C P^1$ sigma model in $1 + 2$ dimensions.
After carefully reviewing the commutative results of Ward and Ruback,
the noncommutative K\"ahler potential is expanded in powers of dimensionless
moduli. In two special cases we sum the perturbative series to analytic
expressions. For any nonzero value of the noncommutativity parameter,
the logarithmic singularity of the commutative metric is expelled
from the origin of the moduli space and possibly altogether.}


\Classification{46L55; 81R60; 81T75}

\renewcommand{\thefootnote}{\arabic{footnote}}
\setcounter{footnote}{0}

\section{Introduction and summary}

The $\C P^1$ sigma model in $1+2$ dimensions is a paradigm for soliton
studies~\cite{zakr,mantonbook}. In particular, it provides the simplest
example for a nontrivial dynamics of slowly-moving lumps, following the
adiabatic approximation scheme of Manton~\cite{manton}. In a slice of the
charge-two sector, the moduli-space metric was worked out and the geodesic
motion was analyzed by Ward~\cite{ward}. The corresponding K\"ahler potential
was then given by Ruback~\cite{ruback} (see also~\cite{dunajski}).

In the case just mentioned, the (restricted) moduli space of static
charge-two solutions is complex two-dimensional and contains ring-like
as well as two-lump conf\/igurations. On the complex line where the lump
size shrinks to zero, the metric develops a logarithmic singularity.
Such divergencies can often be regulated by subjecting the system to
a noncommutative deformation, which introduces a dimensionful deformation
parameter~$\th$. To explore this possibility, we analyze
the Moyal-deformed $\C P^1$ model~\cite{lee} in this paper.

In fact, the (restricted) moduli-space metric for the charge-two sector
of this noncommutative model was already investigated in~\cite{furuta}.
There, the authors show that the metric in question is f\/lat for
$\th \to \infty$ (corresponding to vanishing values of the dimensionless
moduli) and possesses a~smooth $\th \to 0$ limit (which is attained for
inf\/inite values of the dimensionless moduli).
However, these f\/indings do not establish the removal of the logarithmic
singularity for f\/inite values of~$\th$ or amount to an explicit computation
of the K\"ahler potential.

In this paper, we review the commutative results and present a power-series
expansion of the deformed K\"ahler potential in the `ring' regime of the
moduli space. For the f\/irst time, this is achieved for arbitrary values
of~$\th$. We verify the commutative limit and sum up the perturbation series
on the would-be singular line in the `two-lump' domain via the Gel'fand--Yaglom
method. There is a curious connection with the eigenvalues of the spheroidal
wave equation. Around the origin of the moduli space, the K\"ahler potential
is shown to be analytic, which substantiates the claim of~\cite{furuta}.
Perturbative expressions for the moduli-space metric follow
via dif\/ferentiation, and the two-lump scattering behavior may be quantif\/ied.

\section[The $\C P^1$ model and its solitons]{The $\boldsymbol{\C P^1}$ model and its solitons}

The $\C P^1$ or, equivalently, the O(3) sigma model describes the dynamics
of maps from $\R^{1,2}$ with a~metric $(\eta_{\m\n})=\textrm{diag}(-1,+1,+1)$
into $\C P^1\simeq\frac{{\rm SU}(2)}{{\rm U}(1)}\simeq S^2$.
There are various ways to parametrize the target space, for instance by
hermitian rank-one projectors~$P$ in~$\C^2$,
\[
P=P^\+= P^2 =T (\Tdag T)^{-1}\Tdag ,
\]
or else by vectors~$T\in\C^2$ modulo complex scale,
\[
T= \Big(\begin{smallmatrix} p \\[4pt] q \end{smallmatrix}\Big)
 \sim \Big(\begin{smallmatrix} u \\[4pt] 1 \end{smallmatrix}\Big)
\qquad\textrm{with}\quad u= \frac{p}{q} ,
\]
so that the f\/ield degree of freedom is a single function~$u$ taking values
in the extended complex plane $\dot\C\simeq\C P^1$.

Introducing coordinates on $R^{1,2}$,
\[
(x^\mu)  = (t,x,y) \qquad\textrm{with}\quad \mu=0,1,2 \qquad \mbox{and}\qquad
z=x+\im y  ,
\]
we can formulate the action as
\begin{gather*}
S =-4\int \diff^3x\;\tr\,\eta^{\m\n}\pa_\m P\,\pa_\n P  =-4\int \diff^3x\;(\Tdag T)^{-1}\eta^{\m\n}\pa_\m\Tdag(\unity-P)\,\pa_\n T
\\
\phantom{S}=-4\int \diff^3x\;(1+\bar uu)^{-2}\eta^{\m\n}\pa_\m\bar u\,\pa_\n u  ,
\end{gather*}
where $\bar{u}$ is the complex conjugate of~$u$, and only the last equality
uses the commutativity of the functions.
For later convenience, we also def\/ine the kinetic and potential energy density,
\begin{gather*}
{\cal T} =4 (\Tdag T)^{-1}\dot\Tdag (1-P) \dot T
 =\frac{4 \dot{\bar u} \dot u}{(1+\bar uu)^2} \qquad\mbox{and} \\
{\cal V}
 =8 (\Tdag T)^{-1}\pa_\zb\Tdag(1-P) \pa_zT+(\pa_z\leftrightarrow\pa_\zb)
 =\frac{8 \pa_\zb\bar u \pa_zu}{(1+\bar uu)^2}+(\pa_z\leftrightarrow\pa_\zb) ,
\end{gather*}
respectively, so that
\[
S  = \int \diff^3x\;({\cal T}-{\cal V}) ,\qquad \mbox{and}\qquad
E = \int \diff^2z\;({\cal T}+{\cal V})
\]
yields the total energy of the conf\/iguration, which is conserved in time.
Clearly, action and energy are form-invariant under translations and rotations
of the domain~$\R^2$ (at f\/ixed~$t$),
\[
z  \mapsto  z+\la \qquad \mbox{and}\qquad  z  \mapsto  \ep^{\im\mu}z ,
\]
as well as under global SO(3)~rotations of the target,
\[
u \mapsto \frac{au+b}{-\bar{b}u+\bar{a}}
\qquad\textrm{with}\quad \bar aa+\bar bb=1  .
\]

Classically, one is interested in the extrema of~$S$ whose energy is f\/inite.
Among the static conf\/igurations, $\dot u=0$ (hence ${\cal T}=0$),
those are all well known:
\[
\delta S=0 \quad\Rightarrow\quad \delta E=0 \quad\Rightarrow\quad
\delta \int \diff^2z\;{\cal V}=0 \quad\Rightarrow\quad
u=u(z) \ \ \textrm{or} \ \ u=u(\zb),
\]
with $u$ being a rational function (of $z$ or $\zb$) to ensure f\/inite energy.
Each rational analytic (or anti-analytic) function~$u=\frac{p}{q}$ is a
soliton (or anti-soliton) with a topological charge given by its degree~$n$
(or~$-n$) and with energy $E=8\pi|n|$.
Hence, the soliton moduli space~${\cal M}_n$ for charge~$n$
has complex dimension~$2n+1$. Some of these moduli, however,
correspond to isometries of the domain or the target.

In this paper, we shall investigate only charge-one and charge-two solitons.
Let us characterize their static moduli spaces.
By employing the SO(3) target rotations, in the numerator~$p$ we remove
the highest monomial and restrict the coef\/f\/icient of the second-highest one
to be real and non-negative. This is also true for the third-highest one by
means of a domain rotation. Furthermore, by a common rescaling of $p$ and $q$
we set the coef\/f\/icient of the highest monomial in the denominator~$q$ to unity.
Finally, the domain translation isometry allows us to remove the second-highest
monomial of~$q$, which corresponds to picking a center-of-mass frame for our
conf\/iguration. These choices f\/ix all isometries except possibly for special
values of the remaining moduli. Of course, the full moduli space is recovered
by acting with all isometries. In the charge-one case, we thus get
\[
T(z) =\left(\begin{matrix} \b \\ z \end{matrix}\right)
\quad\Rightarrow\quad
{\cal V} =\frac{8 |\b|^2}{(|\b|^2+|z|^2)^2}
=\frac8{|\b|^2} \frac{1}{(1+|z'|^2)^2}
\]
with $\b\in\R_{\ge0}$ and $z=\b z'$.
This is a single lump of height $|\b|^{-2}$ and width of order~$|\b|$.

For charge two, one f\/inds
\begin{gather*}
T(z) =\left(\begin{matrix} \b z+\g \\ z^2+\eps \end{matrix}\right)
\quad\Rightarrow\quad
{\cal V} =\frac{8 |\b z^2+2\g z-\b\eps|^2}{(|\b z+\g|^2+|z^2+\eps|^2)^2} =
\frac8{|\b|^2} \frac{|{z'}^2+2\g'z'-\eps'|^2}{(|z'+\g'|^2+|{z'}^2+\eps'|^2)^2}
\end{gather*}
with $\b,\g\in\R_{\ge0}$ and $\eps\in\C$. In the last expression, we have
introduced dimensionless quantities by the rescaling
\[
z=\b z'  ,\qquad \g=\b^2\g' \qquad \mbox{and}\qquad \eps=\b^2\eps'  ,
\]
ef\/fectively putting $\b=1$. A dif\/ferent situation arises for the
special value $\b=0$.  Here, one can also rotate away the phase of~$\eps$
and should rather use $z=\sqrt{\g} z'$ to arrive at
\[
{\cal V}=\frac{32}{|\g|} \frac{|z'|^2}{\big(1+\big|{z'}^2+\frac{\eps}{\g}\big|^2\big)^2} .
\]
One may check that ${\cal V}$ integrates to $16\pi$ in both cases.
This energy density can take a variety of shapes, depending on the values
of the moduli. Two well-separated lumps appear for $|\eps|>|\b|^2$ and
$|\eps|>|\g|$, while ring-like structures emerge in the regime
$|\g|>|\b|^2$ and $|\g|>|\eps|$.

\section{Moduli space metric}

So far, we have only considered static solutions to the sigma model.
For dynamical issues, we must bring back the time dependence. Rather than
attempting to solve the full equations of motion $\delta S=0$ for $u(t,z,\zb)$,
we resort to the adiabatic approximation valid for slow
motion~\cite{mantonbook},
\[
u(t,z,\zb)  \approx  u(z\,|\,\a(t)),
\]
where $u(z|\a)$ denotes a static soliton depending holomorphically on
moduli parameters~$\a$. For simplicity we suppress here the moduli labels
but let $\a$ represent the holomorphic set~$\{\a\}$.\footnote{We apply isometries to undo possible phase restrictions on $\b$ or $\g$.}
By~allowing these moduli to vary with time, we approximate the
true time-dependent solution by a~sequence of snapshots of static solutions.
In this way, the dynamics in the conf\/iguration space of maps,
$u:\R\to\textrm{maps}(\C,\C P^1)$ via $t\mapsto u(t,\cdot)$, gets projected
to the `mechanics' of a particle moving in the f\/inite-dimensional moduli space
for a f\/ixed topological charge, $\a:\R\to{\cal M}_n$.

Since the potential energy of the soliton conf\/igurations is independent
of~$\a$, the kinetic energy provides an action principle for~$\a(t)$:
the extrema of
\begin{gather*}
\int \diff^3x\;{\cal T}\bigl[u(\cdot\,|\,\a(t))\bigr]
 =4\int \diff{t}\left[ \int\diff^2z\;
(\Tdag T)^{-1}\pa_{\bar\a}\Tdag(1-P)\,\pa_\a T \right]
\dot{\bar\a}\,\dot\a \\
\hphantom{\int \diff^3x\;{\cal T}\bigl[u(\cdot\,|\,\a(t))\bigr]}{}
=4\int \diff{t}\left[ \int\diff^2z\;
\frac{\pa_{\bar\a}\bar u\,\pa_\a u}{(1+\bar uu)^2}\right]
\dot{\bar\a}\,\dot\a
  =:  \frac12\int \diff{t}\;g_{\bar\a\a}(\a)\,\dot{\bar\a}\,\dot\a
\end{gather*}
are just geodesics in~${\cal M}_n$
endowed with the induced K\"ahler metric
\[
g_{\bar\a\a}  = \pa_{\bar\a}\pa_\a{\cal K}  ,
\]
where
\[
{\cal K} =8\int \diff^2z\;\ln \Tdag T =8\int \diff^2z\;\ln(1+\bar uu)
\]
computes the K\"ahler potential from the static soliton
conf\/igurations~$u=u(z|\a)$.
We remark that the freedom of rescaling~$T$ reappears in the ambiguity of
${\cal K}$ due to K\"ahler transformations,
${\cal K}\sim{\cal K}+f(\a)+g(\bar\a)$,
and so we may also use the more divergent formal expression
\[
{\cal K} =8\int \diff^2z\;\ln (\bar pp+\bar qq) .
\]

It turns out that $g_{\bar\a\a}$ diverges for the modulus~$\b$
(and also for the removed $z^n$ coef\/f\/icient in~$p$).
Hence, these particular moduli carry inf\/inite inertia and do not participate
in the dynamics, because changing their values requires an inf\/inite amount
of energy. Consequently, they get degraded to external parameters which are
to be dialled by hand. In the charge-one case, no dynamics remains, which is
consistent with the picture of a single lump sitting in its rest frame.
Nevertheless, it is instructive to reinstate the translation moduli and
verify the f\/lat moduli space. With
$T=\bigl(\begin{smallmatrix} \b \\ z+\de \end{smallmatrix}\bigr)$
we get
\[
{\cal K} =8\int \diff^2z\;\ln\left(1+\frac{|\b|^2}{|z+\de|^2}\right)  .
\]
This is formally independent of~$\delta$ (by shifting $z\mapsto z-\de$)
but it is logarithmically divergent, so we better compute its second
derivatives
\begin{gather*}
\pa_{\bar\b}\pa_\b{\cal K} =8\int \diff^2z\;\frac{|z|^2}{(|\b|^2+|z|^2)^2}
 =\infty \qquad \mbox{and}\qquad
\pa_{\bar\de}\pa_\de{\cal K} =8\int \diff^2z\;\frac{|\b|^2}{(|\b|^2+|z|^2)^2}
 =8\pi  ,
\end{gather*}
as well as $\pa_{\bar\de}\pa_\b{\cal K}=-8\pi\de/\b$.
Hence, we indeed get ${\cal K}=8\pi\bar\de\de$.
Since the center-of-mass motion decouples from the remaining dynamics,
we shall suppress it from now on.

For charge two, the K\"ahler potential ${\cal K}$ reads
\begin{gather*}
8|\b|^2\int \diff^2z'\;\ln\left(1+\frac{|z'+\g'|^2}{|{z'}^2+\eps'|^2}\right)
\qquad\textrm{or}\qquad
8|\g|\int \diff^2z'\;\ln\left(1+\frac{1}{|{z'}^2+\frac\eps\g|^2}\right)
\end{gather*}
depending on whether $\b$ is chosen nonzero or not.
In the f\/irst case, ${\cal K}$ is again divergent, and its derivatives are
not elementary integrable. For the sake of simplicity, we therefore restrict
ourselves to the second (special) case and put $\b=0$ from now on.
The form of the relevant integral reveals that ${\cal K}$ is a function only of
$|\g|$ and $|\eps|$ which, up to an overall dimensional factor, depends merely
on their ratio. The last integral can in fact be executed to yield\footnote{Apart from the normalization, we dif\/fer from~\cite{ruback} by the absence
of a term linear in~$|\eps|$.}
\begin{gather*}
{\cal K} =
16\pi |\g|\int_0^{\pi/2} \diff\th\;\sqrt{1+\left|\frac\eps\g\right|^2\sin^2\th} =
16\pi |\g| E\left(-\left|\frac\eps\g\right|^2\right) =
16\pi r |\cos\vp| E\big(-\tan^2\vp\big)  ,
\end{gather*}
where $E(m=k^2)$ denotes the complete elliptic integral of the second kind
as a function of its parameter $m$ ($k$ is called the elliptic modulus),
and we have parametrized\footnote{There is some ambiguity in the range of the angles. We take $\vp\in[0,2\pi)$.}
\[
\eps =r \ep^{\im\omega}\sin\vp \qquad \mbox{and}\qquad \g =r \ep^{\im\chi}\cos\vp   .
\]
In the $(|\eps|,|\g|)$ plane, the K\"ahler potential grows linearly with
the distance from the origin, with a slope varying between $8\pi^2$
(for $\eps=0$) and $16\pi$ (for $\g=0$). It is continuous but not smooth
on the complex line $\g=0$ ($\vp=\frac\pi2$), which is the localization
locus in the two-lump region because the lump width is of order
$\frac{|\g|}{\sqrt{|\eps|}}$ at a lump separation of order~$2\sqrt{|\eps|}$.

To investigate the two extreme situations, $|\eps|\ll|\g|$ and $|\g|\ll|\eps|$,
we expand the K\"ahler potential in $|\frac\eps\g|$ and $|\frac\g\eps|$,
respectively. For the `ring' regime, $|\eps|\ll|\g|$, we have
\begin{gather}
{\cal K} =8\pi^2|\g| \sum_{\ell=0}^\infty \frac{(-1)^\ell}{1-2\ell}
\left[\frac{(2\ell-1)!!}{(2\ell)!!}\right]^2\left|\frac\eps\g\right|^{2\ell}\nonumber\\
\phantom{{\cal K}}{} =
8\pi^2|\g| \left\{ 1+ \frac14\left|\frac\eps\g\right|^2
-\frac3{64}\left|\frac\eps\g\right|^4 +\frac5{256}\left|\frac\eps\g\right|^6
+\cdots \right\}  ,\label{ring}
\end{gather}
while in the `two-lump' domain, $|\g|\ll|\eps|$, we encounter logarithms,
\begin{gather}
{\cal K} = 16\pi |\eps| \left\{ 1- \sum_{\ell=0}^\infty
\frac{(-1)^\ell}{4(\ell+1)}\left[\frac{(2\ell-1)!!}{(2\ell)!!}\right]^2
c_\ell \left|\frac\g\eps\right|^{2(\ell+1)}\right\} \nonumber\\
\phantom{{\cal K} =}{}  -
4\pi\left|\frac{\g^2}{\eps}\right|\ln\bigl|\frac\g{4\eps}\bigr|^2 \;
{}_2F_1\!\left(\frac12,\frac12;2;-\left|\frac\g\eps\right|^2\right) \nonumber \\
\phantom{{\cal K} }{} = 16\pi |\eps|\left\{ 1- \sum_{\ell=0}^\infty
\frac{(-1)^\ell}{4(\ell+1)}\left[\frac{(2\ell-1)!!}{(2\ell)!!}\right]^2
\left(c_\ell+\ln\left|\frac\g{4\eps}\right|^2\right)
\left|\frac\g\eps\right|^{2(\ell+1)}\right\}  \nonumber\\
\phantom{{\cal K} }{}
 = 16\pi |\eps|\left\{ 1-
\frac14\left(-1+\ln\left|\frac\g{4\eps}\right|^2\right)
\left|\frac\g\eps\right|^2+
\frac1{32}\left(\frac32+\ln\left|\frac\g{4\eps}\right|^2\right)
\left|\frac\g\eps\right|^4\right.\nonumber\\
\left.\phantom{{\cal K}= }{} -
\frac3{256}\left(2+\ln\left|\frac\g{4\eps}\right|^2\right)
\left|\frac\g\eps\right|^6+\cdots\right\} \label{lumps}
\end{gather}
with
\begin{gather*}
c_0=-1 ,\qquad c_1=\frac32  ,\qquad c_2=2  ,\\
c_\ell=\frac2\ell+\frac3{\ell+1}+\frac4{\ell+2}+\frac4{\ell+3}
+\cdots+\frac4{2\ell-1} \quad\textrm{for}\quad \ell\ge3.
\end{gather*}

The metric coef\/f\/icients
\begin{gather*}
g_{\bar\g\g}=\int\frac{8 |z^2+\eps|^2  \diff^2z}{(|\g|^2+|z^2+\eps|^2)^2}  ,
\qquad\!\!
g_{\bar\eps\eps}=\int\frac{8 |\g|^2 \diff^2z}{(|\g|^2+|z^2+\eps|^2)^2}  ,
\qquad\!\!
g_{\bar\g\eps}=\int\frac{-8 \g\bar\eps  \diff^2z}{(|\g|^2+|z^2+\eps|^2)^2}
\end{gather*}
may of course be expressed in terms of complete elliptic integrals~\cite{ward}.
The geodesic motion in this metric cannot be found in closed form,
except for special motions~$\a(t)$,
\begin{gather*}
\dot\omega=\dot\chi=\dot\vp=0 \quad\Rightarrow\quad
(\diff s)^2  =\frac{4\pi}r |\cos\vp| E\big(-\tan^2\vp\big) (\diff r)^2  ,
\end{gather*}
which yields $r(t)=r_0+h(\vp) t^2$ with a specif\/ic function~$h(\vp)$.

\section{Moyal deformation}

The task of this paper is the Moyal deformation of the Ward metric and the
K\"ahler potential presented in the previous section.
One way to describe such a noncommutative deformation of the $z\zb$~plane
is by giving the following `quantization rule':
\[
\textrm{coordinates}\quad(z,\zb)\quad\longmapsto\quad\textrm{operators}\quad
(Z,\bar{Z})\quad\textrm{with}\quad [Z,\bar{Z}]=2\th=\textrm{const},
\]
where these operators may be realized as inf\/inite matrices
\begin{gather*}
Z=\sqrt{2\th} a=\sqrt{2\th}\left( \begin{smallmatrix}
0 & 0 &&& \\  \sqrt{1} & 0 & 0 && \\  & \sqrt{2} & 0 & 0 & \\[-6pt]
&& \sqrt{3} & 0 & \ddots \\[-6pt] &&& \ddots & \ddots \end{smallmatrix} \right), \qquad 
\bar{Z}=\sqrt{2\th} a^\+=\sqrt{2\th}\left( \begin{smallmatrix}
0\,& \sqrt{1}&&& \\  0 & 0 & \sqrt{2}&& \\  & 0 & 0 & \sqrt{3}&\\[-6pt]
&& 0 & 0 & \ddots \\[-6pt] &&& \ddots & \ddots \end{smallmatrix} \right)
.
\end{gather*}
More generally,
\[
\textrm{functions}\quad f(z,\zb)\quad\longmapsto\quad\textrm{operators}\quad
F=f(Z,\bar{Z})\big|_\textrm{sym},
\]
where `sym' indicates a symmetric ordering of all monomials in $(Z,\bar{Z})$.
Naturally, derivatives turn into inner derivations,
\[
\pa_z\quad\mapsto\quad
\frac1{2\th}[\,\cdot\,,\bar{Z}]=\frac1{\sqrt{2\th}}[\,\cdot\,,a^\+] \qquad \mbox{and}\qquad
\pa_\zb\quad\mapsto\quad
\frac1{2\th}[Z,\,\cdot\,]=\frac1{\sqrt{2\th}}[a,\,\cdot\,],
\]
and the integral over the complex plane becomes a trace over the
operator algebra,
\[
\int\diff^2z\;f(z,\zb) \quad\longmapsto\quad 2\pi\th\,\tr\,F.
\]
A highest-weight representation space~${\cal F}$ for the Heisenberg algebra,
$[a,a^\+]=1$, is easily constructed from a vacuum~$|0\>$,
\[
a\,|0\>=0 \quad\Rightarrow\quad {\cal F}=\textrm{span}
\left\{|n\>=\frac1{\sqrt{n!}}(a^\+)^n|0\>\ |\ n=0,1,2,\ldots\right\},
\]
where the basis states are the normalized eigenstates of the `number operator'
$N=a^\+a$,
\begin{equation} \label{basis}
N\,|n\>=n\,|n\> \qquad \mbox{and}\qquad  \<n|n\>=1 \qquad\textrm{for}\quad n=0,1,2,\ldots.
\end{equation}

The Moyal-deformed $\C P^1$ model is def\/ined by copying most def\/initions
of the previous section, but taking the entries of $P$ and $T$ to be
operator-valued. Since, in this context, $q$ may not have an inverse,
we avoid using~$u$ as a variable and work with $p$ and $q$ instead.
Because the deformation has traded functions on the $xy$~plane with
operators on~${\cal F}$, densities such as~${\cal T}$ or~${\cal V}$ are
less intuitive objects, but may still be visualized via the Moyal--Weyl map.
The noncommutative solitons are found by taking $T$ to be polynomial in~$a$,
i.e.~both $p$ and $q$ are polynomial of degree~$n$, and their moduli
are identical to the commutative ones\footnote{In addition to these `non-Abelian' solitons, which smoothly deform the
standard commutative solitons, there exist a plethora of `Abelian' solitons,
which are singular in the commutative limit~\cite{lepo,chu,klawunn}.}.
It is important to note that the deformation has introduced a new dimensionful
parameter, $\th$. Therefore, we may relate all dimensional quantities to~$\th$
and pass to dimensionless parameters,
\[
Z=\sqrt{2\th} a,\qquad \b=\sqrt{2\th} b,\qquad \de=\sqrt{2\th} d,\qquad
\g=2\th g,\qquad \eps=2\th e.
\]
As a consequence, ${\cal K}$ and $g_{\bar\a\a}$ will depend on all moduli
individually and not only on their ratios. Of course, in the commutative
limit~$\th\to0$, the ratios will again dominate.

As a warm-up, let us reconsider the charge-one soliton (with $b$ frozen but
including the translational moduli~$d$), now given by
\[
T=\left(\begin{matrix} b \\ a+d \end{matrix}\right)
\quad\Rightarrow\quad
{\cal K} = 16\pi\th\,\tr\,\ln \Tdag T =
16\pi\th\,\tr\,\ln\bigl( \bar bb+\big(a^\+ + \bar d\big)(a + d) \bigr)  .
\]
Since by a unitary basis change in $\cal F$ we can shift $a\mapsto a - d$,
this expression is again formally independent of $d$, but it is divergent:
\begin{equation} \label{ncK1}
\frac{\cal K}{16\pi\th}=
\sum_{n=0}^\infty\<n|\ln(\bar bb+N)|n\> =
\sum_{n=0}^\infty\ln(\bar bb+n) =
-\ln\Gamma(\bar bb)+\lambda \bar bb+\mu  ,
\end{equation}
where the divergence is hidden in the ambiguous coef\/f\/icients
$\lambda$ and~$\mu$, which may depend on $b$ and~$\bar b$.
To f\/ix this ambiguity, we f\/irst take derivatives and then shift away
the $d$~dependence:
\begin{gather*}
\frac{g_{\bar dd}}{16\pi\th}=\frac{\pa_{\bar d}\pa_d{\cal K}}{16\pi\th}
 =\tr \Bigl[
(\bar bb + N)^{-1} \bigl(1-a (\bar bb + N)^{-1}a^\+\bigr)\Bigr]\\
\hphantom{\frac{g_{\bar dd}}{16\pi\th}=\frac{\pa_{\bar d}\pa_d{\cal K}}{16\pi\th}}{}
=\tr \Bigl[
(\bar bb + N)^{-1}\bigl(1-(N + 1)(\bar bb + N + 1)^{-1}\bigr)\Bigr]\\
\hphantom{\frac{g_{\bar dd}}{16\pi\th}=\frac{\pa_{\bar d}\pa_d{\cal K}}{16\pi\th}}{}
 =\bar bb \, \tr \Bigl[(\bar bb + N)^{-1}(\bar bb + N + 1)^{-1}\Bigr]
=\sum_{n\ge0}\frac{\bar bb}{(n+\bar bb)(n+1+\bar bb)} = 1  ,
\end{gather*}
while $g_{\bar bb}$ is still inf\/inite. Hence, $\lambda$ remains arbitrary
but $\mu=\bar dd$ up to irrelevant terms. With~$b$ f\/ixed, we therefore get
${\cal K}=16\pi\th\bar dd=8\pi\bar\de\de$,
the same f\/lat metric as in the commutative case.
Note that even though the modulus~$b$ has inf\/inite inertia, it is needed
to regulate the K\"ahler potential~(\ref{ncK1}),
which blows up at $b =0$.\footnote{In the operator formalism, the inversion of operators is sometimes complicated
due to zero modes, but may still be accomplished by means of partial isometries
or Murray--von Neumann transformations (see, e.g.~\cite{furuuchi}).}

\section{Deformed rings}

We now turn to the nontrivial charge-two case with the choice of $\b=0$,
def\/ined by
\begin{gather*} 
 T=\left(\begin{matrix} g \\ a^2+e \end{matrix}\right)
\quad\Rightarrow\quad
{\cal K}= 16\pi\th\,\tr\,\ln \Tdag T \qquad \textrm{with}\\
 \Tdag T = \bar gg+\big(a^{\+2}+\bar e\big)\big(a^2+e\big)
= \bar gg+N(N-1) + e a^{\+2} + \bar e a^2+\bar ee .
\end{gather*}
Of course, ${\cal K}$ is divergent, but the singularity is removable,
and $\pa_{\bar gg}{\cal K}$ and $\pa_{\bar e}\pa_e{\cal K}$ already converge.
Like the modulus~$b$ in the previous section, here~$g$ plays the role of
a regulator, but this time its inertia is f\/inite.
This expression is not amenable to exact analytic computation,
but we can attempt to establish power series expansions
in $\bar ee$ or in~$\bar gg$.
In this section, we investigate the `ring' regime $|e|\ll|g|$.

It is easy to set up an expansion around $e=0$, since $\Tdag T|_{e=0}$
is already diagonal in our basis~(\ref{basis}).
Note that no zero-mode issue arises since $\bar gg>0$. Writing
\[
\Tdag T= G + E \quad\textrm{with}\quad
G = \bar gg+N(N-1) \qquad\textrm{and}\qquad
E = e a^{\+2} + \bar e a^2 + \bar ee
  =: \; \nwarrow + \swarrow + \leftarrow  ,
\]
the Taylor series of $\ln(1+x)$ unfolds to
\begin{gather*} 
 \frac{\cal K}{16\pi\th} =\tr\,\ln(G+E) = \tr\,\ln G   -
\sum_{k=1}^\infty\frac{(-1)^k}{k}\,\tr\,\bigl(G^{-1} E\bigr)^k\\
\hphantom{\frac{\cal K}{16\pi\th}}{} = \tr\ln G + \bar ee\,\tr\gi -\frac22\bar ee\,\tr\gi\up\gi\dn
-\frac12(\bar ee)^2\tr\gi\gi
+\frac33(\bar ee)^2\tr\gi\gi\up\gi\dn\\
\hphantom{\frac{\cal K}{16\pi\th}=}{}
+\frac33(\bar ee)^2\tr\gi\up\gi\gi\dn
+\frac13(\bar ee)^3\tr\gi\gi\gi
-\frac24(\bar ee)^2\tr\gi\up\gi\dn\gi\up\gi\dn \\
\hphantom{\frac{\cal K}{16\pi\th}=}{}
-\frac44(\bar ee)^2\tr\gi\up\gi\up\gi\dn\gi\dn
-\frac44(\bar ee)^3\tr\gi\gi\up\gi\gi\dn \\
\hphantom{\frac{\cal K}{16\pi\th}=}{}
-\frac44(\bar ee)^3\tr\gi\gi\gi\up\gi\dn
-\frac44(\bar ee)^3\tr\gi\up\gi\gi\gi\dn
-\frac14(\bar ee)^4\tr\gi\gi\gi\gi \\
\hphantom{\frac{\cal K}{16\pi\th}=}{}
+\frac55(\bar ee)^3\tr\gi\gi\up\gi\dn\gi\up\gi\dn
+\frac55(\bar ee)^3\tr\gi\up\gi\gi\dn\gi\up\gi\dn \\
\hphantom{\frac{\cal K}{16\pi\th}=}{}
+\frac55(\bar ee)^3\tr\gi\gi\up\gi\up\gi\dn\gi\dn
+\frac55(\bar ee)^3\tr\gi\up\gi\gi\up\gi\dn\gi\dn \\
\hphantom{\frac{\cal K}{16\pi\th}=}{}
+\frac55(\bar ee)^3\tr\gi\up\gi\up\gi\gi\dn\gi\dn
+\frac55(\bar ee)^3\tr\gi\up\gi\up\gi\dn\gi\gi\dn \\
\hphantom{\frac{\cal K}{16\pi\th}=}{}
+\frac55(\bar ee)^4\tr\gi\gi\gi\gi\up\gi\dn
+\frac55(\bar ee)^4\tr\gi\gi\gi\up\gi\gi\dn \\
\hphantom{\frac{\cal K}{16\pi\th}=}{}
+\frac55(\bar ee)^4\tr\gi\gi\up\gi\gi\gi\dn
+\frac55(\bar ee)^4\tr\gi\up\gi\gi\gi\gi\dn
+\frac15(\bar ee)^5\tr\gi\gi\gi\gi\gi \\
\hphantom{\frac{\cal K}{16\pi\th}=}{}
-\frac26(\bar ee)^3\tr\gi\up\gi\dn\gi\up\gi\dn\gi\up\gi\dn
-\frac66(\bar ee)^3\tr\gi\up\gi\up\gi\up\gi\dn\gi\dn\gi\dn \\
\hphantom{\frac{\cal K}{16\pi\th}=}{}
-\frac66(\bar ee)^3\tr\gi\up\gi\up\gi\dn\gi\up\gi\dn\gi\dn
-\frac66(\bar ee)^3\tr\gi\up\gi\dn\gi\up\gi\up\gi\dn\gi\dn \\
\hphantom{\frac{\cal K}{16\pi\th}=}{}
+\ O\big((\bar ee)^4\big)   ,
\end{gather*}
displaying all terms to order $(G^{-1} E)^6$ and $(\bar ee)^3$.
One sees that for a given power~$k$ of $G^{-1} E$, there is a sum over
all cyclic paths of length~$k$, where each step is either~$\nwarrow$ or
$\swarrow$ or~$\leftarrow$, separated by a factor of~$\gi$. All terms
containing~$\leftarrow$ can be resummed into the shift operator
$\exp(\bar ee\,\pa_{\bar gg})$, which shortens the above to
\begin{gather*}
\frac{\cal K}{16\pi\th} = \exp\bigl(\bar ee \pa_{\bar gg}\bigr)
\biggl\{ \tr\ln G - \bar ee\,\tr\gi\up\gi\dn
-\frac12(\bar ee)^2\tr\gi\up\gi\dn\gi\up\gi\dn\\
\hphantom{\frac{\cal K}{16\pi\th} =}{}
-(\bar ee)^2\tr\gi\up\gi\up\gi\dn\gi\dn
-\frac13(\bar ee)^3\tr\gi\up\gi\dn\gi\up\gi\dn\gi\up\gi\dn\\
\hphantom{\frac{\cal K}{16\pi\th} =}{}
-(\bar ee)^3\tr\gi\up\gi\up\gi\up\gi\dn\gi\dn\gi\dn
-(\bar ee)^3\tr\gi\up\gi\up\gi\dn\gi\up\gi\dn\gi\dn\\
\hphantom{\frac{\cal K}{16\pi\th} =}{}
-(\bar ee)^3\tr\gi\up\gi\dn\gi\up\gi\up\gi\dn\gi\dn +\cdots\biggr\} \\
\hphantom{\frac{\cal K}{16\pi\th}}{}
 = \exp\bigl(\bar ee\,\pa_{\bar gg}\bigr)\,\tr \biggl\{
\ln G - \bar ee\,\nwarrow\swarrow -  (\bar ee)^2 \biggl(\frac12
 \nwarrow\swarrow\nwarrow\swarrow
+\nwarrow\nwarrow\swarrow\swarrow
\biggr) - (\bar ee)^3  \\
\hphantom{\frac{\cal K}{16\pi\th}=}{}
\times \biggl(\frac13
 \nwarrow\swarrow\nwarrow\swarrow\nwarrow\swarrow
+\nwarrow\nwarrow\nwarrow\swarrow\swarrow\swarrow
+\nwarrow\nwarrow\swarrow\nwarrow\swarrow\swarrow
+\nwarrow\swarrow\nwarrow\nwarrow\swarrow\swarrow
\biggr) +\cdots\biggr\}   .
\end{gather*}
Using
\begin{gather*}
\up|n\>=\sqrt{(n+1)(n+2)}|n+2\>  ,\qquad
\dn|n\>=\sqrt{n(n-1)}|n-2\> \qquad\textrm{and}\\
\gi|n\>=\frac1{\bar gg+n(n-1)}|n\>  ,
\end{gather*}
the above traces convert into inf\/inite sums of rational functions of~$n$.
After repeated partial fraction decomposition these sums can be evaluated to
\begin{gather}
 \frac{\cal K}{16\pi\th}=
\ln\bar gg  +  \ln\cos W   +
\bar ee \pi^2\frac{\bar gg}{4\bar gg + 3} \frac{\tan W}{W}
\nonumber \\
\phantom{\frac{\cal K}{16\pi\th}=}{}
+ (\bar ee)^2\pi^4\biggl\{
\frac{48(\bar gg)^4+200(\bar gg)^3-33(\bar gg)^2+27\bar gg}
     {4 (4\bar gg+3)^3(4\bar gg+15)} \frac{\tan W}{W^3}  -
\frac{(\bar gg)^2}{2\,(4\bar gg + 3)^2} \frac{\sec^2W}{W^2}\biggr\}
\nonumber \\
\phantom{\frac{\cal K}{16\pi\th}=}{}
 + (\bar ee)^3\pi^6\biggl\{
\frac{1}{8 (4\bar gg+3)^5(4\bar gg+15)^2(4\bar gg+35)}
\big(10240(\bar gg)^8+171520(\bar gg)^7\nonumber\\
\phantom{\frac{\cal K}{16\pi\th}=}{}
+878336(\bar gg)^6+1161920(\bar gg)^5-
       354936(\bar gg)^4+549414(\bar gg)^3-13770(\bar gg)^2\nonumber\\
\phantom{\frac{\cal K}{16\pi\th}=}{}
       +6075\bar gg\big)
            \frac{\tan W}{W^5}
  -
\frac{48(\bar gg)^5+200(\bar gg)^4-33(\bar gg)^3+27(\bar gg)^2}
      {4 (4\bar gg+3)^4(4\bar gg+15)} \frac{\sec^2W}{W^4}  \nonumber\\
\phantom{\frac{\cal K}{16\pi\th}=}{}
      +
\frac{(\bar gg)^3}{3 (4\bar gg+3)^3} \frac{\tan W\sec^2W)}{W^3}
\biggr\}
 +\ O\big((\bar ee)^4\big)\label{ncring}
\end{gather}
with the def\/inition
\[
W = \frac\pi2\sqrt{1-4\bar gg}   .
\]
The leading term was determined via
\[
\pa_{\bar gg}\tr\ln G = \tr\,G^{-1} = \sum_{n\ge0} \frac{1}{\bar gg+n(n-1)}
= \frac1{\bar gg}  +  \frac{\pi^2}{2} \frac{\tan W}{W}  ,
\]
and a constant as well as the $\ln\bar gg$ term in ${\cal K}$ may be omitted.

To each order in $\bar ee$, the expression~(\ref{ncring}) is exact in~$\bar gg$
and, hence, valid for arbitrary values of~$\th$. For strong noncommutativity,
when $g\to0$ but $|\frac{e}{g}|\ll1$ f\/ixed,
potential poles due to $\tan W\sim\sec W\sim(\pi\bar gg)^{-1}$ are
always compensated by suitable powers of~$\bar gg$.
To check our computation, let us take the opposite, commutative limit,
\[
\th\ \to\ 0 \quad\textrm{with}\ \ \g, \ \eps \ \textrm{f\/ixed}
\quad\Rightarrow\quad g,\ e\ \to\ \infty \quad\textrm{with} \ \
\frac{e}{g}=\frac{\eps}{\g}\ \ \textrm{f\/ixed} .
\]
For $g\to\infty$, the expansion~(\ref{ncring}) takes the form
\begin{gather*}
\frac{\cal K}{16\pi\th} =\ln\bar gg  -  \ln2 +  \pi\sqrt{\bar gg} \biggl\{
\left(1-\frac1{8\bar gg}+\cdots\right)
+  \left|\frac{e}{g}\right|^2
\left(\frac14-\frac5{32\bar gg}
+  \cdots\right)
\\
\hphantom{\frac{\cal K}{16\pi\th} =}{}
+\left|\frac{e}{g}\right|^4
\left(-\frac{3}{64}+\frac{35}{512\bar gg}+\cdots\right)
+\left|\frac{e}{g}\right|^6
\left(\frac{5}{256}-\frac{105}{2048\bar gg}+\cdots\right)
+  \cdots\biggr\}\\
\hphantom{\frac{\cal K}{16\pi\th}  }{}
  \simeq  \frac{\pi|\g|}{2\th} \biggl\{ 1  +
\frac14\left|\frac{\eps}{\g}\right|^2  -
\frac{3}{64}\left|\frac{\eps}{\g}\right|^4 +
\frac{5}{256}\left|\frac{\eps}{\g}\right|^6  -
\frac{175}{16384}\left|\frac{\eps}{\g}\right|^8  +
\cdots \biggr\}  + O(\th) ,
\end{gather*}
after dropping the irrelevant logarithmic and constant terms
through `$\simeq$'. Indeed, the leading contributions reproduce the commutative
K\"ahler potential~(\ref{ring}) in the $|\eps|\ll|\g|$ `ring' regime.

\section{Deformed lumps}

More interesting however is the $|\g|\ll|\eps|$ `two-lump' domain,
which at $\th=0$ featured a weak logarithmic singularity for $\g\to0$,
where the two lumps are localized inf\/initely sharply.
To analyze this situation, we need to expand~${\cal K}$ around~$g=0$,
in powers and perhaps also logarithms of~$\bar gg$, generalizing~(\ref{lumps})
to f\/inite values of~$\th$. To this end, we are interested in the eigenvalues of
\begin{gather*} 
\Tdag T\big|_{g=0} = (a^{\+2}+\bar e)(a^2+e)
= N(N-1) + e a^{\+2} + \bar e a^2+\bar ee  =:  F   .
\end{gather*}

Representing the noncommutative coordinates on $L_2(\R)\ni f:\R\to\R$,
\begin{gather*}
a=\frac1{\sqrt{2}}(x+\pa_x)=
\frac1{\sqrt{2}}\ep^{-x^2/2}\pa_x \ep^{x^2/2} \qquad \mbox{and}\\
a^\+=\frac1{\sqrt{2}}(x-\pa_x)=
\frac1{\sqrt{2}}\ep^{x^2/2}\pa_x \ep^{-x^2/2} ,
\end{gather*}
one gets
\begin{gather*}
\bigl[F f\bigr](x)  \equiv  \left[
\frac14\pa_x^4-x \pa_x^3+\left(x^2 - \frac12+\frac{e+\bar e}{2}\right) \pa_x^2
-2 e x \pa_x+(2e x^2 - e+\bar ee)\right] f(x) \\
\phantom{\bigl[F f\bigr](x)}{}
= \la(e) f(x)  ,
\end{gather*}
which via Fourier transformation and change of variables is equivalent to
\begin{gather} \label{EVproblem}
\left[-\pa_z(1-z^2) \pa_z+\frac{1}{1-z^2}+\bar ee\big(1-z^2\big)\right]f(z)
=\la(e)\,f(z) \qquad\textrm{with}\quad z\in[-1,1].
\end{gather}
This equation matches with the one def\/ining spheroidal (scalar) wave functions
\cite{abramowitz,li},
\begin{gather*}
\left[-\pa_z\big(1-z^2\big) \pa_z+\frac{m^2}{1-z^2}-\g^2\big(1-z^2\big)\right]
f_{mn}(z) =\la_{mn}(\g) f_{mn}(z)   ,
\qquad\textrm{with}\quad m\in\Z
\end{gather*}
and $n=0,1,2,\ldots$ (in our convention) counting the discrete spheroidal
eigenvalues~$\la_{mn}(\g)$.
Clearly, we have $m=1$ and $\g^2=-\bar ee$ (the oblate case),
hence  $\la_n(e)=\la_{1n}(\im|e|)$.
For small values of $\bar ee$ one f\/inds the expansion~\cite{abramowitz}
\begin{gather*}
\la_n(e)  = n(n-1)  \biggl\{ 1 + \frac{2}{(2n-3)(2n+1)}\bar ee \\
\phantom{\la_n(e)  =}{} +
\frac{2(4n^4-8n^3-35n^2+39n+63)}{(2n-5)(2n-3)^3(2n+1)^3(2n+3)}
(\bar ee)^2 + \cdots \biggr\}
  =:  n(n-1) \tilde{\la}_n(e) ,
\end{gather*}
where the two zero modes of~$F$, namely $\la_0=\la_1=0$, are explicit.
Therefore, we may write
\begin{gather}
\frac{\cal K}{16\pi\th} =\sum_{n=0}^\infty\ln\bigl[\la_n(e)+\bar gg\bigr]=
2\ln\bar gg  +  \sum_{n=2}^\infty\ln\bigl[\la_n(e)+\bar gg\bigr] \nonumber\\
\hphantom{\frac{\cal K}{16\pi\th}}{}
=2\ln\bar gg  + \sum_{n=2}^\infty\ln\bigl[n(n-1)+\bar gg\bigr]  +
\sum_{n=2}^\infty\ln\left[1 +
\frac{n(n-1)}{n(n-1)+\bar gg} (\tilde\la_n(e)-1)\right] .\label{sphK}
\end{gather}
The role of $g$ as a regulator is obvious; the f\/irst term carries the $F$
zero modes. After expanding the logarithm under the last sum, one can
perform the sums and nicely reproduces all terms in~(\ref{ncring}).

We have not found an asymptotic expansion of the spheroidal eigenvalues
around $|e|=\infty$, and so it is dif\/f\/icult to analyze the `two-lump' domain
in general. For a f\/irst impression, let us expand (\ref{ncring}) in powers
of $\bar gg$ and collect the $e$~dependence of each term:
\begin{gather}
\frac{\cal K}{16\pi\th}= 2\ln\bar gg  +   \left\{
\ln\pi+\frac23\bar ee-\frac4{45}(\bar ee)^2
+\frac{64}{2835}(\bar ee)^3-\frac{32}{4725}(\bar ee)^4+\cdots\right\}
\nonumber\\
\hphantom{\frac{\cal K}{16\pi\th}=}{}
+ \bar gg   \left\{ 1-\frac29\bar ee+\frac{56}{675}(\bar ee)^2
-\frac{656}{19845}(\bar ee)^3+\frac{1216}{91125}(\bar ee)^4+\cdots\right\}
\nonumber\\
\hphantom{\frac{\cal K}{16\pi\th}=}{}
+ (\bar gg)^2 \left\{ \left(\frac32-\frac{\pi^2}6\right)
+\left(\frac{62}{27}-\frac{2\pi^2}{9}\right)\bar ee
+\left(\frac{3742}{3375}-\frac{16\pi^2}{135}\right)(\bar ee)^2\right.\label{nclumps}\\
\left. \hphantom{\frac{\cal K}{16\pi\th}=}{}
+\left(\frac{3822944}{10418625}-\frac{32\pi^2}{945}\right)(\bar ee)^3+\cdots\right\}
+(\bar gg)^3 \left\{ \left(\frac{10}3-\frac{\pi^2}3\right)
+\left(\frac{292}{81}-\frac{10\pi^2}{27}\right)\bar ee\right.\nonumber\\
\left.
\hphantom{\frac{\cal K}{16\pi\th}=}{}
+\left(\frac{254846}{151875}-\frac{112\pi^2}{675}\right)(\bar ee)^2
+\left(\frac{1235859892}{3281866875}-\frac{12176\pi^2}{297675}\right)(\bar ee)^3
+\cdots\!\right\}\!
  + O\bigl((\bar g g)^4\bigr) .\!\nonumber
\end{gather}
The $g\to0$ singularity due to the two zero modes is visible in the f\/irst term,
but it is inconsequential in a K\"ahler potential. Besides this, the expression
is devoid of the commutative logarithmic small-$g$ singularity\footnote{The linear piece,
${\cal K}_{\textrm{lin}}=16\pi\th \big(\bar gg+\frac23\bar ee\big)$,
has already been found in~\cite{furuta}. Furthermore,
$\sum\limits_{n=2}^\infty\ln n(n-1)\buildrel{\textrm{reg}}\over=\ln\pi$.}!
From the pattern in~(\ref{ncring}) it is clear that this feature persists
to all orders in the expansion. Apparently, the Moyal deformation has smoothed
out the K\"ahler potential near $\g=\eps=0$, where the two lumps collide.

To attain the analog of~(\ref{ncring}) for the `two-lump' domain, one would
have to sum the series in each pair of curly brackets in~(\ref{nclumps}).
This can actually be achieved for the f\/irst of these series
(the $g$-independent contribution), as we shall demonstrate shortly.
So let us concentrate on the dangerous $g=0$ line from now on.
Because $\tilde\la_n(0)=1$ and
\[
\frac{1}{16\pi\th}{\cal K}\big|_{e=0} = 2\ln\bar gg+\ln\pi+O(\bar gg) ,
\]
at $g=0$ we may subtract this from the K\"ahler potential, and
(\ref{sphK}) simplif\/ies to
\[
\ln\frac{\det (a^{\+2}+\bar e)(a^2+e)}{\det a^{\+2} a^2}
= \lim_{g\to0}\frac{{\cal K}-{\cal K}|_{e=0}}{16\pi\th}
= \sum_{n=2}^\infty\ln\tilde\la_n(e) = -\ln\tilde\la_1(e)  .
\]
The last equality is an observation we have checked to $O((\bar ee)^8)$
but do not know its origin\footnote{It amounts to $\prod\limits_{n=1}^\infty\tilde\la_n(e)=
\prod\limits_{n=1}^\infty\frac{\la_n(e)}{\la_n(0)}=1$,
i.e.~the formal product of all spheroidal eigenvalues is $e$~independent.}.
It allows us to easily push the $\bar ee$ expansion to higher orders,
\begin{gather}
\lim_{g\to0}\frac{{\cal K}-{\cal K}|_{e=0}}{16\pi\th}
= \frac23 \bar ee  -  \frac4{45}(\bar ee)^2  +  \frac{64}{2835}(\bar ee)^3
  -  \frac{32}{4725}(\bar ee)^4  +  \frac{1024}{467775}(\bar ee)^5\nonumber\\
\hphantom{\lim_{g\to0}\frac{{\cal K}-{\cal K}|_{e=0}}{16\pi\th} =}{}
 - \frac{1415168}{1915538625}(\bar ee)^6
 + \frac{32768}{127702575}(\bar ee)^7
 - \frac{14815232}{162820783125}(\bar ee)^8 +\cdots  .\label{higher}
\end{gather}

To identify the function behind this power series, we exploit the
Gel'fand--Yaglom theorem~\cite{gelfand,levit,kirsten}.
Let us go back to the eigenvalue problem~(\ref{EVproblem}) and
stretch the interval $[-1,1]$ to $\R$ by the change of variables
$z=\tanh y$, so that it becomes
\[
\bigl[F(e) f\bigr](y) \equiv
\left[-\cosh^2 y \pa_y^2 + \cosh^2 y + \frac{\bar ee}{\cosh^2 y}\right]
f(y) = \la(e) f(y) .
\]
The Gel'fand--Yaglom theorem states that
\begin{gather} \label{gy}
\frac{\det (a^{\+2}+\bar e)(a^2+e)}{\det a^{\+2} a^2}   \equiv
\frac{\det F(e)}{\det F(0)} =
\lim_{L\to\infty}\frac{\Phi(L)}{\Psi(L)} ,
\end{gather}
where the functions $\Phi(y)$ and $\Psi(y)$ satisfy the following set of
equations and boundary conditions,
\begin{alignat}{4}
& \bigl[F(e) \Phi\bigr](y) = 0  , \qquad &&
\Phi (-L) = 0 , \qquad &&
\Phi'(-L) = 1 ,& \nonumber\\
& \bigl[F(0) \Psi\bigr](y)  = 0 , \qquad &&
\Psi (-L) = 0 , \qquad &&
\Psi'(-L) = 1 .& \label{gycond}
\end{alignat}
respectively. The solution can be given analytically:
\begin{gather*}
\Phi(y) = \frac1{\sqrt{\bar ee}}\sinh\bigl[
\sqrt{\bar ee} (\tanh y + \tanh L)\bigr] \cosh y \cosh L\\
\phantom{\Phi(y)}{}
= \frac1{\sqrt{\bar ee}}\sinh\left[
\sqrt{\bar ee} \frac{\sinh(y+L)}{\cosh y \cosh L}\right] \cosh y \cosh L ,
\\
\Psi(y) = \sinh (y + L)  ,
\end{gather*}
which leads to
\[
\frac{\Phi(L)}{\Psi(L)} =
\frac{\sinh(2\sqrt{\bar ee}\tanh L)}{2\sqrt{\bar ee}\tanh L}
\quad\buildrel{L\to\infty}\over\longrightarrow\quad
\frac{\sinh(2\sqrt{\bar ee})}{2\sqrt{\bar ee}} =
\frac{\det F(e)}{\det F(0)}  .
\]
Hence, we f\/inally arrive at
\[
\lim_{g\to0}\frac{{\cal K}-{\cal K}|_{e=0}}{16\pi\th} =
\ln \frac{\sinh(2\sqrt{\bar ee})}{2\sqrt{\bar ee}} =
\ln\left\{1+\sum_{\ell=1}^\infty\frac{(4\,\bar ee)^\ell}{(2\ell+1)!}\right\}
 ,
\]
whose expansion indeed reproduces all terms in~(\ref{higher}).
This simple expression represents the full noncommutative K\"ahler potential
at $g=0$ and provides an analytic formula for~$\tilde\la_1(e)$.
Moreover, it has the correct commutative limit $e\to\infty$,
\begin{gather*}
\lim_{g\to0}\frac{{\cal K}-{\cal K}|_{e=0}}{16\pi\th}
= 2|e| - \ln|e| - 2\ln 2 + O(\ep^{-4|e|})
\quad\buildrel{2e=\eps/\th}\over\Longrightarrow\quad
\lim_{\th\to0}{\cal K}(\g=0) = 16\pi|\eps|  ,
\end{gather*}
again up to irrelevant constant and $\ln\bar\eps\eps$ terms.
It is tempting to add the constant shift by $\bar gg$ in~(\ref{gy})
and apply the Gel'fand--Yaglom technique for all values of~$g$.
However, we have not been able to solve~(\ref{gycond}) with this shift.

For completeness, we display the metric coef\/f\/icients,
\begin{gather*}
g_{\bar\g\g} =\frac{4\pi}{\th}
\pa_{\bar gg}\left(\bar gg \pa_{\bar gg}\frac{\cal K}{16\pi\th}\right)
=\frac{4\pi}{\th}\,\tr\bigl\{(\Tdag T)^{-1}-\bar gg (\Tdag T)^{-2}\bigr\}
 ,\\
g_{\bar\eps\eps} =\frac{4\pi}{\th}
\pa_{\bar ee}\left(\bar ee \pa_{\bar ee}\frac{\cal K}{16\pi\th}\right)
 =\frac{4\pi}{\th}\,\tr\bigl\{(\Tdag T)^{-1}-
(\Tdag T)^{-1}(\up+\bar e)(\Tdag T)^{-1}(\dn+e)\bigr\}
 ,\\
g_{\bar\g\eps} =\frac{4\pi}{\th}
 g\bar{e} \pa_{\bar gg}\pa_{\bar ee}\frac{\cal K}{16\pi\th}
 =-\frac{4\pi}{\th} g\,\tr\bigl\{(\Tdag T)^{-2}(\up+\bar e)\bigr\} .
\end{gather*}
All these traces converge and should be f\/inite in the entire $\g\eps$~plane.
The coef\/f\/icient~$g_{\bar\eps\eps}$ may be read of\/f~(\ref{ncring}) by replacing
$(\bar ee)^k$ with $k^2(\bar ee)^{k-1}$ in the series. For the other two,
one has to work out the derivatives.

\begin{figure}[t]
\centerline{\includegraphics[width=7cm]{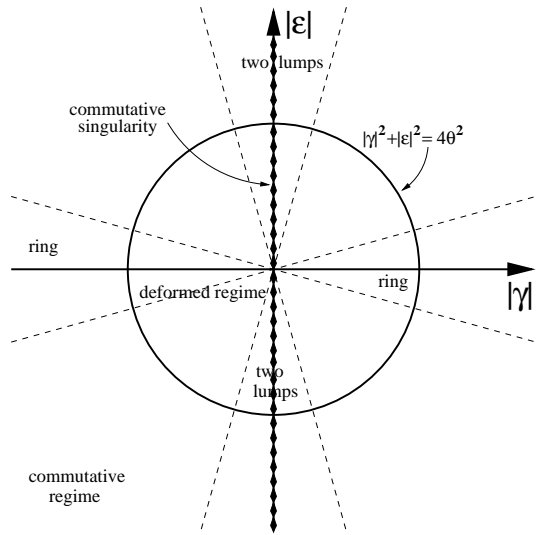}}
\caption{Modulus-of-moduli space.}\label{Fig1}
\end{figure}

\section{Conclusions}

We have investigated the charge-two moduli-space metric in the noncommutative
$\C P^1$ sigma model in $2+1$ dimensions. After decoupling the center of
mass and a convenient dialling of frozen moduli, we f\/ind that the K\"ahler
potential depends only on the combinations $\bar gg$ and $\bar ee$ of the
dynamical complex-valued dimensionless moduli $g$ and~$e$. The noncommutativity
strength~$\sqrt{\theta}$ sets the single scale of the system. In the limit
$|e|\ll|g|\to\infty$, where the solitonic energy density has a ring-like
prof\/ile, our power series in $\bar ee$ matches with the known commutative
K\"ahler potential, which depends only on the ratio $\frac{|e|}{|g|}$.
In the complementary regime $|g|\ll|e|$, where the conf\/iguration splits
into two lumps, we observe that the logarithmic singularity of the commutative
K\"ahler potential is smoothed out by the deformation, which pushes it to
the $\theta=0$ boundary of the moduli space. The $(|\gamma|,|\epsilon|)=2\theta (|g|,|e|)$ plane
     is depicted in Fig.~\ref{Fig1}.

We have expanded the K\"ahler potential to order $(\bar ee)^4$ and to any order
in $\bar gg$, but an analytic expression remains a challenge, which amounts to
computing the spectrum of the spheroidal wave equation for $m=1$ but any~$e$.
However, at $g=0$ we only needed the lowest (regularized) eigenvalue, and
the $\bar ee$ series could be summed to an analytic function via the
Gel'fand--Yaglom trick.

\subsection*{Acknowledgements}

The authors acknowledge f\/inancial support from DAAD, Kennzif\/fer A/08/03929.
O.L.\ is grateful for discussions with N.~Dragon, H.~Grosse, W.~Nahm, C.~Nash
and E.~Schrohe. In particular he thanks A.~Fischer for discovering the relation
to spheroidal wave functions and M.~Rubey for support with his program FriCAS
\cite{fricas}.

\pdfbookmark[1]{References}{ref}
\LastPageEnding

\end{document}